\documentstyle[aps,prl,twocolumn]{revtex}
\begin{document}
\title{Frustrated H-Induced Instability of Mo\,(110)}
\author{Bernd Kohler, Paolo Ruggerone, Steffen Wilke, and Matthias
Scheffler}
\address{Fritz-Haber-Institut der Max-Planck-Gesellschaft, Faradayweg 4-6,
D-14\,195 Berlin-Dahlem, Germany}
\date{submitted 25 May 1994}
\maketitle
\begin{abstract}
Using helium atom scattering Hulpke and L\"udecke recently observed a giant
phonon anomaly  for the
hydrogen covered W\,(110) and Mo\,(110) surfaces.
An explanation which is able to account for this and other experiments is
still lacking.
Below we
present density-functional theory calculations of the atomic and electronic
structure of the clean
and hydrogen-covered Mo\,(110) surfaces. For the full adsorbate monolayer
the
calculations provide evidence for a strong Fermi surface nesting
instability. This explains the observed anomalies and resolves the
apparent
inconsistencies of different experiments.
\end{abstract}
PACS numbers: 68.35.-p,73.20.-r,73.20.Mf\\[0.1cm]
Recently, giant anomalies in the surface phonon spectra have been detected
by helium atom scattering (HAS) on the (110)\,surfaces of tungsten and
molybdenum, when these are covered with a full monolayer of
hydrogen~\cite{HUL92,HUL93,LUE94}.
One critical wave vector ${\bf Q}_{\rm c1}^{\rm exp}$ is along the  [001]
direction (${\overline{\Gamma {\rm H}}}$)  and has a length of 0.95
\AA$^{-1}$ for W and 0.90 \AA$^{-1}$ for Mo.
Anomalies have also been found for directions deviating from
${\overline{\Gamma {\rm H}}}$.
Along ${\overline{\Gamma {\rm S}}}$
they are located at ${\overline{{\rm S}}}$ with a critical wave
vector ${\bf Q}_{\rm c2}^{\rm exp}$ ($|{\bf Q}_{\rm c2}^{\rm exp}|$ = 1.225~\AA
$^{-1}$ for Mo and 1.218~\AA$^{-1}$ for W)~\cite{NOTA1}.
The experimental dispersion curves (see Ref.~\cite{LUE94}) are only available
for ${\overline{\Gamma {\rm H}}}$ and ${\overline{\Gamma {\rm S}}}$.
Thus, we focus our attention particularily on these two directions.
For both surfaces two different softenings
in the surface phonon branches were observed at the critical wave vectors:
a smaller dip
and a very deep and sharp indentation from $\hbar \omega \approx 15$~meV
to $\hbar \omega \approx 2$~meV.
Only the phonon dispersion curves of quasi-one-dimensional conductors,
like KCP [K$_{2}$Pt(CN)$_{4}$Br$_{3}$]~\cite{REN75} are characterized by
similarly deep anomalies strongly localized in reciprocal space.
A pronounced but less sharp damping of  longitudinal surface phonons
was observed for the
(100) surfaces of W~\cite{ERN,schweizer89} and Mo~\cite{mo_phonons} and
was explained in terms
of the nested structure of the Fermi surface~\cite{SMI90,chung92}.
In the case of W\,(100) this Fermi surface nesting
is apparently so strong that it induces a rebonding at the surface and a
structural rearrangement, i.e. a c($2 \times 2$)~surface
reconstruction at temperatures below
$T_{\rm c}\approx 250$~K (for a review see~\cite{jupille94}).
For the clean (110) surfaces of W and Mo no indication of an
anomaly is found, but  this only appears when the surfaces are covered
with a full layer of hydrogen.
Nevertheless, a close link between the surface phonon
anomalies and H vibrations seems to be ruled out, since the
HAS spectra remain practically unchanged when
deuterium is adsorbed instead of hydrogen~\cite{HUL93,LUE94}.

Angular resolved photoemission (ARP) studies have been performed by Kevan
and coworkers~\cite{JEO88,JEO89,GAY89} for clean and hydrogen
covered W\,(110) and Mo\,(110). These studies give no evidence of
the existence of parallel segments of the Fermi-surface contours separated
by wave vectors comparable with the HAS determined
critical wave vectors and thus there appeared
to be compelling
reason for abandoning the nesting mechanism as a possible origin
of the anomalies seen in HAS.

The situation became even more puzzling when very careful high resolution
electron energy loss
spectroscopy (HREELS) studies~\cite{BAL93} of hydrogen covered W\,(110)
observed the Rayleigh wave phonon branch with the small dip at
${\bf Q}_{\rm c1}^{\rm exp}$ exactly as the HAS results. However, the  giant
anomaly was not detected.

At present there exists no explanation which is able
to account for the observed two phonon anomalies and is
compatible with the different experimental data (HAS, HREELS, and ARP).
We therefore performed
{\em ab-initio} calculations of the surface atomic geometries and the
electronic properties
of the
clean and H-covered Mo\,(110) surface.
The results for the adsorbate covered surface provide clear evidence for
a
Fermi surface nesting instability at wave vectors which are in remarkable
agreement with those at which the phonon anomalies occur for this system
[${\bf Q}_{\rm c1}^{\rm th}= (0.86 \pm 0.02, 0)$~\AA$^{-1}$,
${\bf Q}_{\rm c1}^{\rm exp}=(0.90, 0)$~\AA$^{-1}$, ${\bf Q}_{\rm c2}^{\rm th}=
(1.00 \pm 0.02, 0.71\pm 0.02 )$~\AA$^{-1}$,
${\bf Q}_{\rm c2}^{\rm exp}=(1.00, 0.707)$~\AA$^{-1}$]. We choose
the $x$-axis along [001] and the $y$-axis is along [${\overline 1}10$].
The corresponding directions in reciprocal space are
$\overline{\Gamma{\rm H}}$ and $\overline{\Gamma{\rm N}}$, respectively.
The calculations were performed using density-functional theory
together with the local-density approximation for the
exchange-correlation energy functional~\cite{CEP80}. The
full-potential linearized augmented plane wave
method~\cite{BLA85} is employed which we enhanced by
the calculation of forces~\cite{kohler94}. This enables
an efficient evaluation of the multi-layer substrate relaxation and
the atomic positions at the surface.
The substrate is
modeled by a seven layer slab repeated periodically with
a separation of 8.8~\AA~ of vacuum. The ${\bf {k}}$-integrations
are evaluated on a mesh of 64 equally spaced points in the surface
Brillouin zone (SBZ). The muffin-tin radii are chosen to be
1.27~\AA~and 0.48~\AA~for Mo and H,
respectively. The kinetic-energy cutoff for the plane wave basis needed for
the interstitial region is set to  12~Ry, and the $(l,m)$ representation
(inside the muffin tins) is taken up to $l_{\rm max}= 8$.
For the potential expansion we use a plane-wave cutoff energy
of 64~Ry~and a $(l,m)$ representation with $l_{\rm max}= 4$.
The core states as well as the valence states are treated
non-relativistically. The calculated in-plane lattice
constant is 3.13~\AA~without including zero point vibrations,
which compares well with the measured bulk lattice parameter
(3.148~\AA~at room temperature~\cite{kata79}).
The values of the surface relaxation
for the clean surface are in excellent agreement  with
those calculated by Methfessel {\it et al.}~\cite{MET92} who
employed the full-potential linearized muffin-tin orbitals approach.

For a meaningful theoretical analysis of the surface electronic structure
it is important to use the correct atomic geometry. For the hydrogen
adsorption system at a full monolayer, i.e. at the coverage for which the
phonon anomaly was observed experimentally, neither the hydrogen adsorption
site nor the substrate relaxation have been determined so far.
In Table \ref{tab:relax} we summarize the calculated adsorption energy
differences $\Delta E_{\rm ad}$ as well
as the relaxation parameters for all possibly important hydrogen sites.
The geometries are defined in Fig.~\ref{geometry}.
The results identify the
hollow site as the most stable position. The [$\bar{1}$10]-offset
from the long-bridge position is $y_{\rm H}= 0.55 \pm 0.01$~\AA.
Within the computational error the hydrogen relaxes exactly in
the geometric threefold site and there is no theoretical
evidence for a pronounced top-layer-shift reconstruction in
agreement with the low-energy electron diffraction experiments
\cite{alt87}. In comparison both bridge positions are clearly
energetically unfavorable, and the on-top position is even worse.
The theory predicts
that the clean-surface inward relaxation of $-4.5~\%d_{0}$,
where $d_{0}$ is the bulk inter-layer spacing,
is reduced for the long-bridge, short-bridge, and hollow geometries of the
adsorbed hydrogen. In the case of the on-top adsorption site we
obtain an outward relaxation.
Note that the numerical error for all calculated
inter-layer spacings is about $\pm 0.3~\%d_0$.

In Fig.~\ref{disper} we show the Fermi surfaces for the
clean Mo\,(110) surface and for the H-covered surface.
On the left the theoretical results of the $N$-particle ground state are
plotted and the right side displays the experimental ARP data of Jeong
{\em et al.}~\cite{JEO89}. For the adsorbate system we present the
results for the hollow position.

The shaded areas are the projection of the
bulk Fermi surface onto the (110) SBZ. For the experimental
analysis it was calculated using a tight-binding interpolation
scheme~\cite{papa86}. Differences between this projection
and the one of the present work may arise
from the lack of self-consistency of the tight-binding approach.
For the clean surface (see Fig.~\ref{disper}a) the calculations identify
four bands, which are highly localized at the surface.
There is a band-circuit centered at $\overline{\Gamma}$, one centered at
$\overline{\rm S}$, and one centered at $\overline{{\rm N}}$.
Furthermore, there is a band more or less parallel to the
$\overline{\Gamma{\rm N}}$ direction. The latter one is of ($d_{3z^{2}-r^{2}},
d_{xz}$)
character, and also the other surface features are due to Mo $d$ bands.
All of these bands have been
observed experimentally, and for all of them the calculated and measured
${\bf k}$-dependence  is in very good agreement (compare
Figs.~\ref{disper}a and b).
Clearly worse agreement is found when we
compare the results for the adsorbate systems (see Fig.~\ref{disper}, lower
panels).

At first we discuss the changes of the theoretical Fermi surface induced
by the hydrogen adsorption (Fig.~\ref{disper}a and c), and then we return
to the experiments.
For the present discussion the most important effect
is that the ($d_{3z^{2}-r^{2}}, d_{xz}$) band is shifted away from the
$\overline{\Gamma{\rm N}}$ line.
A significant fraction is now in the stomach shaped band gap.
This shift is not a consequence of  hybridization between
the ($d_{3z^{2}-r^{2}}, d_{xz}$) states and hydrogen orbitals but it results
because
the adsorption of hydrogen changes the surface potential and shifts the
entire ($d_{3z^{2}-r^{2}}, d_{xz}$) band to lower energies. As this band
disperses upward when
${\bf k}$ increases away from
$\overline{\Gamma}$, the Fermi energy cuts the shifted band at a longer
${\bf k}$-vector.
The hydrogen induced bonding states are found at lower energies (about
5~eV below the Fermi level), and the
antibonding states are at higher energies (about 4~eV above the Fermi
energy).

Figure~\ref{disper}c reveals that the Fermi-surface contours associated
to the ($d_{3z^{2}-r^{2}},d_{xz}$)-like band
run parallel to the $\overline{\Gamma {\rm N}}$ direction
and perpendicular to $\overline{\Gamma {\rm S}}$ in significant
parts of the SBZ giving rise to a quasi-one-dimensional nesting.
The {\bf k}-vectors connecting these bands in different
sections of the Brillouin zone are
${\bf Q}_{\rm c1}^{\rm th}$ and ${\bf Q}_{\rm c2}^{\rm th}$.
We recall that the highest occupied Kohn-Sham level of the self-consistent
$N$-electron calculation  equals the electron
chemical potential, i.e. the Fermi {\em level}.
However, the Kohn-Sham Fermi {\em surface} is (in principle) not an
observable.
The agreement between Fig.~\ref{disper}a and b and the fact that the
wave function character does not change significantly upon hydrogen
adsorption
suggests that the difference between the Kohn-Sham Fermi surface and the
true Born-Oppenheimer Fermi surface is not very important for this system.
We therefore conclude that the nesting instability predicted
by the Fermi surface of Fig.~\ref{disper}c can be trusted, and due to
its one-dimensional nature we expect important consequences~\cite{TOS75}.
Such an instability may induce a static distortion provided that the
electronic energy gain is stronger than  the elastic energy cost needed to
displace the top layer atoms. Because the
(110)\,surface is the closest packed and most stable bcc surface
such a Peierls distortion is somewhat unlikely.
When a static distortion is not realized,
there will be strong dynamical consequences (similar to a dynamical
Jahn-Teller effect in molecules) due to thermally excited electron-hole
pairs with a wavevector equal to the critical wavevectors and their
coupling to surface phonons.

Since for metal surfaces helium atoms scatter at a distance of
about 3-4~\AA~in front of the surface HAS will detect electron
charge density oscillations close to the Fermi level~\cite{KAD92}.
For the system Mo\,(110) at the critical wave vectors these
oscillations are
associated with electron-hole pair excitations involving the
($d_{3z^{2}-r^{2}}, d_{xz}$) band.
We therefore propose that the two bands seen by HAS are due to a
hybridization between lattice vibrations and electron-hole pairs.
Thus these studies represent an experimental manifestation of
a nearly one-dimensional Kohn anomaly. While the small dip is
due to a more phonon like excitation, the sharp and
giant anomaly is due to a more electron-hole like excitation of nested
electron-hole pairs. The anomalies persist in the
phonon dispersion curves  along directions deviating from
$\overline{\Gamma {\rm H}}$~\cite{HUL93,LUE94}.
This result is directly
understandable in terms of the theoretical results of Fig.~\ref{disper}c,
because nesting is present also for the directions in the surface Brillouin
zone out of the [001] azimuth. Because of the form of
the ($d_{3z^{2}-r^{2}}, d_{xz}$) band Fermi-surface contour, the anomalies are
particularly strong along $\overline{\Gamma {\rm S}}$ in agreement with
the available experimental data~\cite{LUE94}. We expect a weakening of the
indentations for directions between $\overline{\Gamma {\rm H}}$ and
$\overline{\Gamma {\rm S}}$ because of the less effective nesting.
A careful reanalysis of the experimental data is in progress~\cite{HUL94}.

Our interpretation of the HAS
results also explains why in HREELS only the
Rayleigh mode with the small dip was observed. This technique
couples to vibrations of the nuclei and is practically insensitive to
electronic charge density oscillations~\cite{KAD92}. Because
the sharp and giant
anomaly has only little vibrational character, it is practically impossible
to excite it by high-energy electrons.

The important question remains why the ARP measurements  for the hydrogen
covered Mo\,(110) and the theoretical Kohn-Sham Fermi surface are so much
different. The results of Fig.~\ref{disper}c are due to photoelectrons
which carry the information about the difference of the $N$ particle ground
state and the $N-1$ particle system with a hole at $|\epsilon_{F}, {\bf
k}\rangle$.
Typically an analysis is done under the assumption that the electronic
system is in its ground state and that vibrational excitations
as well as interactions between electronic and vibrational
excitations can be ignored.
The theoretical results shown in Fig.~\ref{disper}c, which are obtained
under exactly this hypothesis, disclose that for this system the assumption
is unfounded. In fact, the calculations predict
highly localized parallel bands at the Fermi surface with a
one-dimensional nesting vector. As a consequence,
at ${\bf Q}_{\rm c1}$ and ${\bf Q}_{\rm c2}$ many electron-hole pairs
will be thermally excited, and
these will couple to  phonons.
This implies a breakdown of the Born-Oppenheimer approximation.
While the calculations identify a strong nesting,
the ARP experiments investigate a system which has already reacted
to this instability.
When a static distortion is hindered,
ARP will measure
a state with a significant number of low energy excitations,
and in particular at the critical wave vectors the electron and lattice
dynamics
cannot be decoupled.
In fact, we note that the experimental Fermi surface
of the H covered surface seems to exhibit a different translational
symmetry than that  of the $(1 \times 1)$ SBZ (see the band circuits at the
$\overline{\rm S}$ point in Fig.~\ref{disper}d). We cannot rule out, however,
that this is due to inaccuracies in the experimental
data or analysis. We are not aware that for any other surface such a serious
breakdown of the
Born-Oppenheimer approximation (compare Figs.~\ref{disper}c and d)  has
been observed so far.
Therefore, some doubts remain at this point if an alternative  explanation
may exist which can also resolve the differences of the experimental
ARP data and the calculated results. Obviously, the static distortion of
the W\,(100) surface is also induced by a Kohn anomaly, but after the
reconstruction has taken place the Born-Oppenheimer approximation is again
valid.

In conclusion, the calculated Kohn-Sham Fermi surface can explain the
HAS and HREELS experiments. It is interesting that the same
band which is responsible for the pronounced Kohn anomaly identified above
for  Mo\,(110) is in fact also responsible for notable features
of all group VIA materials. In particular we mention the spin-density wave
in bulk Cr
and the phonon anomalies along the [001] direction in bulk W and Mo.
Also the reconstruction
of Mo\,(100) and W\,(100) is induced by the nesting of this band.
What makes the anomaly at the (110) so particularly strong is that the
adsorption of hydrogen shifts it into a region of ${\bf k}$-space where
it becomes localized at the surface. Thus the band becomes two- and the
nesting one-dimensional.
Neither the hydrogen wave functions nor the hydrogen vibration are
{\em directly} involved in the resulting anomalies. The effect of hydrogen is
mainly to change the potential at the surface and to shift the Mo surface
states. Thus, it is well possible, that other adsorbates can have a similar
effect on the surface vibrations and electron-phonon coupling.
We also expect that  for the clean surface a  weak
signature of the ($d_{3z^{2}-r^{2}}, d_{xz}$) band should exist at ${\bf k}
\approx
0.6$~\AA$^{-1}$. However, for the clean surface the band
is not a surface state but a broad and less localized surface resonance.
The thermal excitations of
electron-hole pairs at Mo\,(110):H and the  suggested breakdown of the
Born-Oppenheimer approximation
call for additional photoemission experiments, in which the Fermi surface
is carefully studied also at lower  temperatures.
In fact, also a detailed analysis of the temperature dependence of the
small dip seen in HAS and HREELS would be most interesting.
Furthermore, calculations of the electron-phonon coupling, electron-hole
excitations and frozen phonons at these surfaces would be very
helpful, but for transition metal surfaces this is not yet feasible with
the required accuracy.

We thank J. L\"udecke, E. Hulpke, and E. Tosatti for
stimulating discussions.

\begin{figure}
\caption{Surface geometry of Mo\,(110) with the long-bridge (L),
short-bridge (S), hollow (H) and on-top (T) sites marked. $y_{\rm H}$ is
the $[\bar{1}10]$-offset of the hollow position from the long-bridge site.}
\label{geometry}
\end{figure}
\begin{figure}
\caption{Fermi surfaces of clean Mo\,(110) [upper panels]
and Mo\,(110) with a full
monolayer of hydrogen [lower panels]. The experimental results are from
\protect\cite{JEO89}.
Shaded areas are the projection of the bulk Fermi
surfaces
onto the (110) surface Brillouin zone. The solid (dotted) lines of the
theoretical results denote
surface resonances or surface states which are localized by more than
60 \% (30 \%) in the two top Mo layers.
${\bf Q}_{\rm c1}^{\rm th}$ and
${\bf Q}_{\rm c2}^{\rm th}$ are the theoretical critical wavevectors.}
\label{disper}
\end{figure}
\begin{table}
\begin{tabular}{c|c|cccc}
& clean & hollow & long bridge & short bridge & on top \\
\hline
$d_{\rm H-Mo}$ (\AA)        & ---    &$  1.07$&$  1.08$&$  1.32$&$  1.76$\\
$\Delta d_{12}~(\%d_{0})$   &$ -4.5 $&$ -2.1 $&$ -1.9 $&$ -2.8 $&$ +1.8 $\\
$\Delta d_{23}~(\%d_{0})$   &$ +0.5 $&$ +0.1 $&$ -0.1 $&$ +0.3 $&$ -1.2 $\\
$\Delta d_{34}~(\%d_{0})$   &$  0.0 $&$ -0.1 $&$ -0.3 $&$ -0.3 $&$ +0.1 $\\
$\Delta E_{\rm ad}$ (eV)    & ---    &$  0.0 $&$ -0.23$&$ -0.28$&$ -1.11$\\
\end{tabular}
\caption{Calculated geometries and adsorption-energy differences  for clean
Mo\,(110) and for the H-covered surface. For the latter the results for
the long-bridge, short-bridge, hollow, and on-top sites (see Fig. 1) are
compiled. $d_{\rm H-Mo}$ represents the height of the hydrogen atom above
the surface while $\Delta d_{ij}$ is the percentage change of the
inter-layer distance between the $i$-th and the $j$-th layer with
respect to the bulk inter-layer spacing $d_{0}$.}
\label{tab:relax}
\end{table}
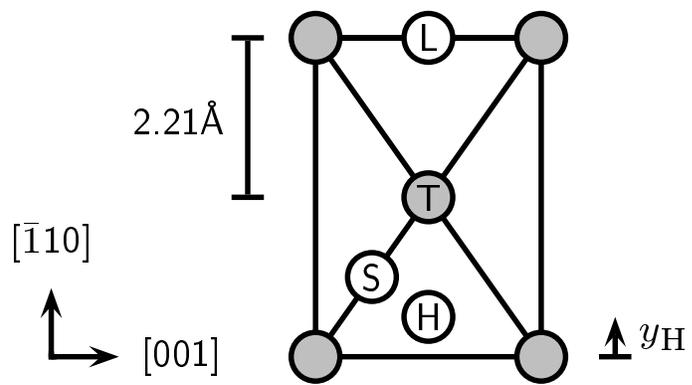
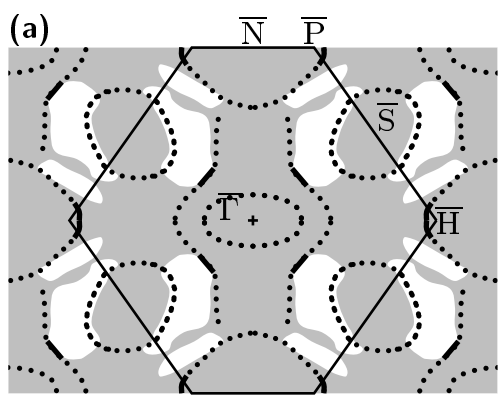
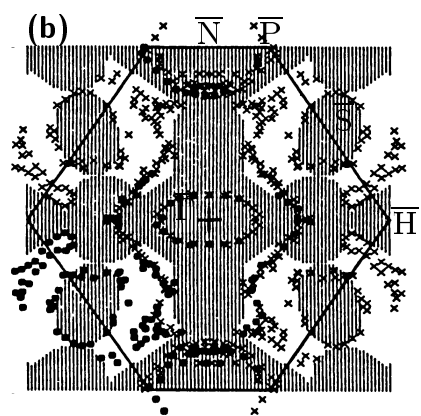
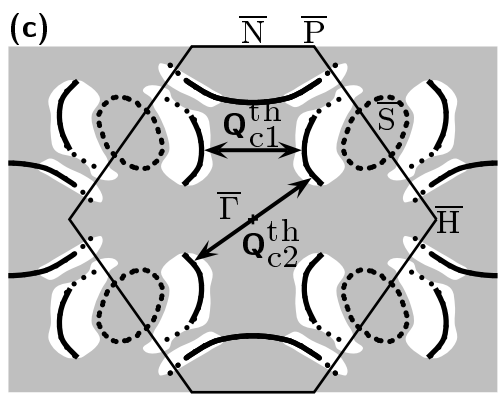
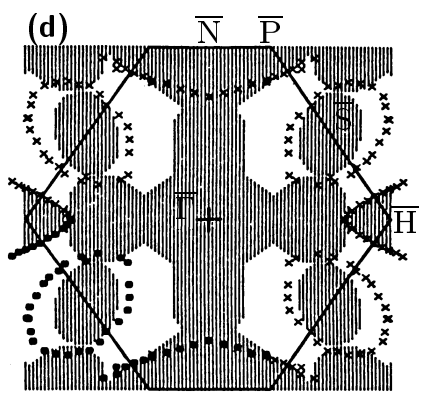
\end{document}